\documentclass{article}
\usepackage{epsfig}
\newcommand{\bfr}{\begin{flushright}}
\newcommand{\efr}{\end{flushright}}
 
\begin{document}
\title{Condensation of Yang--Mills Field at High Temperature 
in the Presence of Fermions
}
\author{ 
Atsushi Nakamula\\
Department of Physics, Tokyo Metropolitan University,\\
Setagaya-ku, Tokyo 158, Japan\\
and\\
Kiyoshi Shiraishi\\
Department of Physics, Faculty of Science,
Ochanomizu University,\\
1-1 Otsuka 2, Bunkyou-ku,
Tokyo 112, Japan
}
\date{acta phys.~slov.~{\bf 42} (1992), No.~6,  pp. 338--343 
}
\maketitle
\begin{abstract}
The possible condensation of the time-component of Yang-Mills field at
finite temperature is  discussed in the presence of Dirac fermions. We
show that the condensation forms regardless of  the number of
fundamental and adjoint fermion species coupled to the Yang-Mills field.
The effect  of finite density of fermions is also investigated and it is
shown that the magnitude of the condensation  is also independent of
the densities.
\end{abstract}

\section{INTRODUCTION}

In the Euclidean formulation of quantum field theory at finite
temperature, the imaginary time variable ($\tau$)  is compactified with a
period $\beta=T^{-1}$, where $T$ is the temperature of the system
\cite{1}.

Then a classical background of gauge field $\langle A_0\rangle\ne 0$
cannot be transformed into $A_0=0$ in general,  because only gauge
transformations which satisfy the periodic boundary condition in the
Euclidean time  direction are permitted \cite{2}.

It has been claimed that a condensation of field $A_0$ in non-Abelian
theory arises at finite temperature \cite{3,4}.
Stimulated by this result, in the present paper we examine the
condensation of the gauge field at finite  temperature in the presence
of fermions. If the phenomenon of gauge field condensation has some
relevance  to the physics of quark-gluon plasma phase \cite{5}, the
presence of fermions as ``quarks'' must be crucial for  thorough
analysis.

As the simplest case, we carry out in this paper the calculations of
the effective potential of finite  temperature $SU(2)$ gauge theory up
to two-loop order with fundamental and adjoint fermions. The
generalization to the $SU(3)$ case will be reported in the future.

\section{EFFECTIVE POTENTIAL}

We show the effective potential, or, free energy in background
space-time $R^d\times S^1$, where $S^1$ stands for the compact time
direction. By global gauge rotation, we can choose a classical
background gauge field in a matrix form as
\begin{equation}
A_0=\frac{T\Phi}{2g}\left(
\begin{array}{cc}
1 & 0 \\
0 & -1
\end{array}
\right)\,,
\end{equation}
where $g$ is the Yang-Mills coupling constant. The value of $A_0$
condensation is described by the expectation value of $\Phi$.

The free energy up to two-loop in the pure $SU(2)$ Yang-Mills theory in
four dimensions has been obtained by many authors, for example in
ref.~\cite{4,6}. In general dimensions we find that free energy up to
two-loop level is expressed as
\begin{eqnarray}
F_g&=&-(D-2)\frac{\Gamma(D/2)}{\pi^{D/2}}T^D\sum_{k=1}^\infty
\frac{2\cos k\Phi +1}{k^D}\nonumber \\
& &+\frac{g^2}{2}\left(\frac{\Gamma(D/2)}{\pi^{D/2}}\right)^2
T^{2D-4}\left(\sum_{k=1}^\infty
\frac{\cos k\Phi}{k^{D-2}}\right)
\left(\sum_{k=1}^\infty
\frac{\cos k+2\Phi}{k^{D-2}}\right)\,,
\end{eqnarray}
where $D=d+1$ is the dimension of the space-time.

The summations of the series in this expression can be reduced to
polynomials provided that $D$ is an even number. Those are known as
Bernoulli polynomials $B_s(x)$ \cite{7} and we can write them as
\begin{equation}
\sum_{k=1}^\infty
\frac{\cos
k\Phi}{k^{2n}}=\frac{(-1)^{n-1}}{2(2n)!}(2\pi)^{2n}B_{2n}(\Phi/2\pi)\,,
\end{equation}
where $0\le x\le 2\pi$ and $n$ is an integer. The Bernoulli polynomials
are, for example, 
\begin{eqnarray}
B_2(x)&=&x^2-x+1/6,	\\
B_4(x)&=&x^4 -2x^3+x^2-1/30\,,
\end{eqnarray}
etc.

By using these polynomials, we can simply expand the effective
potential $F_g$ for $\Phi$ around $\Phi=0$. (Note that at one-loop
level, or equivalently when $g$ is set to zero, global minimum is at
$\Phi=0$.)

Here we set $D=4$, our dimensions. We get an expansion
\begin{equation}
F_g=-\left(\frac{\pi^2}{15}+\frac{g^4}{96}\right)T^4-\frac{g^2}{6\pi}
|\Phi|T^4+\left(\frac{1}{3}+\frac{5g^2}{24\pi^2}\right)\Phi^2T^4
+O(\Phi^3)\,.
\end{equation}
From this, one can find that the minimum of the free energy is located
at
\begin{equation}
\Phi=\frac{g^2}{4\pi}\,,
\end{equation}
in the leading order in $g$. Then a condensate forms and the value is 
$\langle
A_0\rangle=\frac{gT}{8\pi}\left(\begin{array}{cc}
1& 0\\ 0& -1\end{array}\right)$.

We thus conclude that $A_0$ condensation is formed at finite temperature
in the Yang-Mills system with small $g$. This value of condensation is
consistent with $g\ll 1$ and the perturbation is valid.

It is notable that this ``phenomenon'' is peculiar for $D=4$. For $s\ge 
4$, it is known that $B_s(x)$ does not have the linear term in $x$.
Therefore if gauge field condensation occurs in $D>4$, the effect of
higher-order terms of $\Phi$ in the potential must be essential. One can
easily find, however, that it is impossible to yield the condensation by
perturbative way in the case with $D>4$.

An approach to non-perturbative effects on Yang-Mills condensation has
been studied by the present authors in ref.~\cite{8}. We cannot touch on
this immense subject in the present paper, and here we concentrate on the
perturbative argument.

\section{FERMION DIAGRAM}

\begin{figure}[ht]
\begin{center}
\includegraphics[width=3cm]{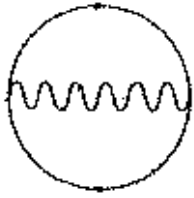}
\caption{A two-loop Feynman diagram for a fermion contribution.}
\label{f1}
\end{center}
\end{figure}

Next, we compute the fermion diagram. The one-loop result is obtained
by the standard technique and is widely known. See for example
ref.~\cite{9}. The two loop diagram in Fig.~\ref{f1}  contributes to the
free energy. We consider Dirac fermions in fundamental and adjoint
representations of $SU(2)$. Calculations are straightforward tasks and
after some rearrangements the results of two-loop contributions to free
energy of the fundamental and adjoint fermions in general dimensions are
represented as follows: 
\begin{eqnarray}
& &F_f(\mbox{two-loop})=-\frac{g^2}{2}2^{[D/2]}\frac{D-2}{2}T^{2D-4}
\nonumber \\
& &\times 
\left[\frac{1}{2}\left\{I\left(\pi+\frac{\Phi}{2}\right)+
I\left(\pi-\frac{\Phi}{2}\right)\right\}\{I(0)+2
I(\Phi)\}\right.\nonumber
\\ & &\left.-I\left(\pi+\frac{\Phi}{2}\right)
I\left(\pi-\frac{\Phi}{2}\right)
-\frac{1}{4}\left\{I\left(\pi+\frac{\Phi}{2}\right)^2+
I\left(\pi-\frac{\Phi}{2}\right)^2\right\}\right]\,,\\ &
&F_a(\mbox{two-loop})=-\frac{g^2}{2}2^{[D/2]}\frac{D-2}{2}T^{2D-4}
\nonumber \\
& &\times 
\Bigl[2\left\{I\left(\pi+{\Phi}\right)+
I\left(\pi-{\Phi}\right)\right\}\{I(0)+I(\Phi)-I(\pi)\}\nonumber
\\ & &-4I\left(\pi\right) I\left({\Phi}\right)
-\left\{I\left(\pi+{\Phi}\right)^2+
I\left(\pi-{\Phi}\right)^2\right\}\Bigr]\,, 
\end{eqnarray}
where $[~]$ in the first line of
each equation is Gauss' symbol and 
\begin{equation}
I(x) =\frac{1}{2\pi^{D/2}}	\Gamma\left(\frac{D-2}{2}\right)
\sum^\infty_{k=1}
\frac{\cos kx}{k^{D-2}}\,.
\end{equation}

Each $I$ in eqs.~(8) and (9) comes from the momentum integration, which
corresponds to a line in the diagram (Fig.~\ref{f1}). There are two
integrations in each two-loop diagram, because of the momentum
conservation. In order to obtain the result of (8) and (9), one needs to
sum up the contributions of the diagrams in which $SU(2)$ suffixes are
assigned to each line. At the moment, one must notice that each
propagator contains the coupling to the background gauge field, i.e.
$\Phi$, and it depends on the $SU(2)$ suffix.

Again we can investigate the location of the minimum by expansion with
respect to small $\Phi$. For $D = 4$, we get the expansion of the free
energy of fermions at one- and two-loop level for each species: 
\begin{eqnarray}
F_f&\approx&-\left(\frac{7\pi^2}{90}+\frac{5g^2}{192}\right)T^4-
\frac{g^2}{24\pi}|\Phi|T^4+\frac{1}{12}\Phi^2T^4+\cdots\,, \\
F_a&\approx&-\left(\frac{7\pi^2}{60}+\frac{5g^2}{48}\right)T^4-
\frac{g^2}{6\pi}|\Phi|T^4+\frac{1}{3}\Phi^2T^4+\cdots\,, 
\end{eqnarray}
where we
discard the $O(g^2\Phi^2)$ term, which is irrelevant for our purpose.
Thus in each case free energy has perturbative minimum; the minimum of
$F_f$ and $F_a$ are both located at $\Phi=g^2/4\pi$ when $g\ll 1$. This
is a marked result. Total free energy is given by $F_g+N_fF_f+N_aF_a$,
where
$N_f$ and
$N_a$ are the number of fermion species which belong to the fundamental
and adjoint representations of $SU(2)$, respectively. Since the total
free energy is given as a linear combination of $F$'s, the magnitude of
the condensate turns out to be independent of the number of the fermion
species.

\section{FINITE DENSITY EFFECT}

Finally, we examine finite density effect of fermions. Finite density
effects of ``quarks'' may have much importance in the physics of a
possible formation of quark-gluon plasma at hadronic collision and very
early universe \cite{5}.

We introduce chemical potentials for fermion numbers. This can be
incorporated in the calculation of diagrams by considering the general
phase of the fermionic field which appears in the boundary condition:
\begin{equation}
\psi(\tau+\beta)=-e^{i\delta}\psi(\tau)\,,	
\end{equation}
and setting $\delta$ to an imaginary value $-i\mu$, where $\mu$ is
identified to the chemical potential.

For $SU(2)$ fundamental and adjoint Dirac fermions the contributions up
to two-loop thermodynamic potential ($\Omega$) including respective
chemical potentials potential of $\Phi$ behave near $\Phi\approx 0$ as
\begin{eqnarray}
\Omega_f&\approx&-\left(\frac{7\pi^2}{90}T^4+\frac{\mu^2T^2}{3}+
\frac{\mu^4}{6\pi^2}\right)+g^2\left(\frac{5}{192}T^4+
\frac{3\mu^2T^2}{32\pi^2}+\frac{3\mu^4}{64\pi^4}\right)\nonumber \\
& &-\frac{g^2}{24\pi}|\Phi|\left(T^4+\frac{3\mu^2T^2}{\pi^2}\right)+
\frac{1}{12}\Phi^2\left(T^4+\frac{3\mu^2T^2}{\pi^2}\right)+\cdots\,,\\
\Omega_a&\approx&-\left(\frac{7\pi^2}{60}T^4+\frac{\mu^2T^2}{3}+
\frac{\mu^4}{4\pi^2}\right)+g^2\left(\frac{5}{48}T^4+
\frac{3\mu^2T^2}{8\pi^2}+\frac{3\mu^4}{16\pi^4}\right)\nonumber \\
& &-\frac{g^2}{6\pi}|\Phi|\left(T^4+\frac{3\mu^2T^2}{\pi^2}\right)+
\frac{1}{3}\Phi^2\left(T^4+\frac{3\mu^2T^2}{\pi^2}\right)+\cdots\,.
\end{eqnarray}

Here we omitted the $O(g^2\Phi^2)$ term again. These expressions are
given by the replacement $\delta\rightarrow -i\mu$ in analogous
calculations to eqs.~(8) and (9), and rewriting them using the Bernoulli
polynomials. The result of calculations including the ``twists'' can be
found in ref.~\cite{8} (note the difference in the definitions of
$\delta$). Judging from these, we find the location of the minimum of
total thermodynamic potential is unchanged, $\Phi=g^2/4\pi$, in the
leading order. It is revealed that the magnitude of the $A_0$
condensation is independent of the density of fermions in the
fundamental and adjoint representations of $SU(2)$ at least in the
perturbative regime.

To summarize all the results, we declare that gauge field condensation
seems to take place even if fermion fields are present. The finite
density of fermions does not affect the magnitude of the condensate. To
obtain the next-leading contribution in $\langle\Phi\rangle$, i.e.,
$O(g^3)$ we have to calculate an infinite sum of diagrams \cite{6}. We
leave the higher-order calculation for a future subject.

In ref.~\cite{4a}, Belyaev claimed that the contribution of ring
diagrams of gluons cancels the linear term of $\Phi$ in the free energy.
Nevertheless, it is yet unknown whether the fermion contribution we have
calculated is canceled by higher-order contribution or not. So far, we
do not know fermionic diagrams which cancel the linear term. We feel the
importance of a further study of the fermionic matter field in the
high-temperature YM theory.

The ``realistic'' case of $SU(3)$ ``quarks'' will be studied by a similar
method with straightforward but tedious calculations. The investigations
of perturbative and non-perturbative approaches and analytical and
non-analytical methods will progress side by side, and we hope to report
the full non-Abelian effects and physical insight in separate
publications.

\section*{ACKNOWLEDGMENTS}

The authors would like to thank S.~Hirenzaki for some useful
comments. One of the authors (KS) would like to thank A.~Sugamoto for
reading this manuscript.
 
KS is indebted to Soryuusi shogakukai for financial support. He would
like to acknowledge financial aid of Iwanami F\=ujukai.


\end{document}